\documentclass[12pt]{article}

\usepackage{amsmath} 
\usepackage{graphics}
\usepackage{graphicx}
\usepackage{epsfig}
\usepackage{indentfirst}
\usepackage{comment}
\usepackage{amssymb}
\usepackage{color}
\usepackage{url}
\usepackage{hyperref}
\usepackage[affil-it]{authblk}
\usepackage[margin=1.2in]{geometry}

\newcommand{\aap}{    {\it Astron. Astrophys.}}

\newcommand{\apj}{    {\it Astrophys. J.}}
\newcommand{\apjl}{   {\it Astrophys. J. Lett.}}

\newcommand{\solphys}{{\it Solar Phys.}}

\newcommand{\icarus}{ {\it Icarus}}
\newcommand{\araa}{   {\it ARA\&A}}

\begin{document}
\title{Helioseismic Imaging of Fast Convective Flows Throughout the Near-Surface Shear Layer}

\author{Benjamin~J.~Greer
  \thanks{Electronic address: \texttt{benjamin.greer@colorado.edu}}}
\author{Bradley~W.~Hindman}
\author{Nicholas~A.~Featherstone}
\author{Juri~Toomre}
  \affil{JILA and Department of Astrophysical and Planetary Sciences, University of Colorado, Boulder CO}

\maketitle




\begin{abstract}
Using a new implementation of ring-diagram helioseismology, we ascertain the strength and spatial scale of convective flows throughout the near-surface shear layer.
Our ring-diagram technique employs highly overlapped analysis regions and an efficient method of 3D inversion to measure convective motions with a resolution that ranges from $3 \ \mathrm{Mm}$ at the surface to $80 \ \mathrm{Mm}$ at the base of the layer. 
We find the rms horizontal flow speed to peak at $427 \ \mathrm{m \ s^{-1}}$ at the photosphere and fall to a minimum of $124 \ \mathrm{m \ s^{-1}}$ between $20 \ \mathrm{Mm}$ and $30 \ \mathrm{Mm}$.
From the velocity amplitude and the dominant horizontal scales seen at each depth, we infer the level of rotational influence on convection to be low near the surface, but transition to a significant level at the base of the near-surface shear layer with a Rossby number varying between 2.2 to as low as 0.1.
\end{abstract}


\section{INTRODUCTION}
\label{sec:intro}

The Sun's differential rotation is maintained by turbulent transport of angular momentum through the action of Reynolds stresses. 
The spatial and temporal correlations between flow components that generate such stresses are thought to be imposed through Coriolis deflection of convective motions. 
Due to this deep connection between convection and rotation, the helioseismic measurement of the rotation rate as a function of latitude and depth (e.g., \cite{thompson_2003}) has provided an inflexible constraint on numerical simulations of the convection zone. 
An even more stringent constraint would be imposed by direct helioseismic determination of the convective flow velocities. 
As the first significant step in this direction, \cite{2012PNAS..10911928H}, hereafter HDS12, have used time-distance helioseismology to infer the subsurface convective flow amplitude. 
Specifically, they found an exceptionally low rms velocity which had an \textit{upper} limit of $1 \ \mathrm{m \ s^{-1}}$ at a depth of $30 \ \mathrm{Mm}$. 
Such low speeds are in conflict with numerical simulations, which generate speeds that are several orders of magnitude larger (\cite{vogler_2005}, \cite{miesch_2008}, \cite{rempel_2009}, \cite{trampedach_2011}, \cite{weber_2011}, \cite{hotta_2014}).
This discrepancy has cast doubt in both directions, implying that either the dynamical balance achieved in many simulations is far from the one that holds in the Sun, or that helioseismic methods (unknowingly) underestimate velocities below the first few megameters in depth. 
In this context, we attempt to provide another source of comparison using a new implementation of ring-diagram helioseismology to directly image convective flows throughout the near-surface shear layer. 
With this technique we find convective amplitudes much larger than those deduced by HDS12, and quite similar to the amplitudes predicted by numerical simulations. 

In Section \ref{sec:methods} we describe our methods, which are largely an advancement of standard ring-diagram helioseismology. 
In Section \ref{sec:results} we present our measurements of the horizontal flow in the near-surface solar interior. 
Finally, in Section \ref{sec:discussion} we compare our findings to both observations and simulations. 

\section{METHODS}
\label{sec:methods}
\subsection{Measurement Procedure}
Ring-diagram helioseismology deduces subsurface flows within the Sun through the measurement of Doppler shifts of the Sun's acoustic wave modes. 
Here, we observe these waves using full-disk, line-of-sight Dopplergrams produced by the \textit{Helioseismic and Magnetic Imager} (HMI) aboard the Solar Dynamics Observatory (SDO). 
We analize 2048 consecutive Dopplergrams with a cadence of 45 seconds beginning on 2010 July 27.
This results in a single analysis period of $25.6 \ \mathrm{hrs}$. 
Portions of each full-disk Dopplergram are projected onto a mosaic of smaller analysis regions called tiles. 
Each square tile is $194 \ \mathrm{Mm}$ in longitude and latitude ($16^{\circ}$ in heliographic angle) and the separation between tile centers in the mosaic is $3 \ \mathrm{Mm}$ ($0.25^{\circ}$) in latitude and longitude. 
All together, the mosaic spans $60^{\circ}$ in longitude and latitude and is comprised of 58,564 individual tiles. 
From image to image in the time series, the central longitude of the projection for each tile is shifted with time at the latitudinally-dependent surface differential rotation rate as measured by \cite{snodgrass_1984}. 
This tracking removes the large-scale velocity signal of solar differential rotation.

Any analysis of velocity measurements is limited in depth by the tile size (larger tiles sample deeper flows) and limited in horizontal resolution by the spacing between adjacent tiles. 
In previous studies the mosaic of tiles has typically had a much sparser spacing, with tile centers separated by half the tile size. 
Thus, standard ring-diagram techniques have achieved only coarse resolution (e.g., $100 \ \mathrm{Mm}$).
Here we decrease the spacing between adjacent tiles while keeping the tile size constant.
Consequently, we are able to substantially refine the resolution through deconvolution without sacrificing depth information. 

After tracking each tile independently through the entire sequence of Dopplergrams, we apply a circular apodization function and create a three-dimensional power spectrum (two horizontal spatial wavenumbers and temporal frequency). 
For each wave mode in the Sun, a sub-surface horizontal flow induces a Doppler shift that is measurable in the spectrum as a frequency splitting. 
We use the Multi-Ridge Fitting (MRF) code (\cite{2014SoPh..289.2823G}) to measure both the Doppler shift and its associated uncertainty for typically $220$ unique modes of different radial order $n$ and horizontal wavenumber $k$.
The measured Doppler shifts exhibit a large-scale systematic bias of currently-unknown origin that depends on disk position and wave mode (\cite{zhao_2012}, \cite{2013ASPC..478..199G}, \cite{kholikov_2014}).
This systematic error has an amplitude of $10$ to $20 \ \mathrm{m \ s^{-1}}$ and contributes only to global scale flows ($\ell < 10$). 
We have measured the systematic using a large set of ring fits spanning 80 days and removed it using the procedure detailed in \cite{2013ASPC..478..199G}. 
The longitudinal mean of each flow component is removed after this step.

\subsection{Inversions}
The Doppler shifts measured by the MRF method are generated by horizontal flows in the solar interior, and adjacent measurements are highly correlated due to our dense tiling scheme. 
Each measured shift is a weighted average over the velocity field within a three-dimensional region. 
The weighting function for this average is called the sensitivity kernel and has a structure that depends on the wave mode used to obtain the Doppler shift (\cite{birch_2007}). 
The broad horizontal and vertical extent of the sensitivity kernels complicate the interpretation of the Doppler shifts as a direct measure of sub-surface flows, since most of the convective structures under consideration are smaller than the kernels. 
However, high-resolution can still be achieved by deconvolving the measurements while accounting for the high degree of correlation.  
We use the Multi-Channel Deconvolution (MCD) procedure from \cite{2012SoPh..276...19J} to achieve high horizontal resolution and produce estimates of the horizontal velocity field at strongly localized points in 3D space. 
This method combines every mode from all 58,564 tiles and finds a balance between producing spatially isolated flow estimates and the final propagated uncertainty. 
The balance is tuned with a regularization parameter, which we choose to vary smoothly with depth such that the signal-to-noise of our results is roughly constant at all depths.
The covariance between any two measurements is assumed to be the product of the uncertainty measured with the MRF method for each mode times the fractional overlap area of the tiles from which the measurements are obtained.

Just as each Doppler shift measurement is related to the velocity field inside the Sun through a sensitivity kernel, each solution point in the inversion is related to the true velocity field through an averaging kernel.
The purpose of the inversion is to produce averaging kernels that are much smaller and spatially localized than the sensitivity kernels. 
For every solution point, we compute the full 3D averaging kernel in order to properly interpret our results. 
The horizontal size of the averaging kernel for a particular depth demonstrates the effective resolution of the inversion solution at that depth. 
The vertical structure of the averaging kernel provides an estimate of not only the vertical resolution, but the actual depth achieved for a given set of modes. 

\begin{figure*}
	\centering
	\includegraphics{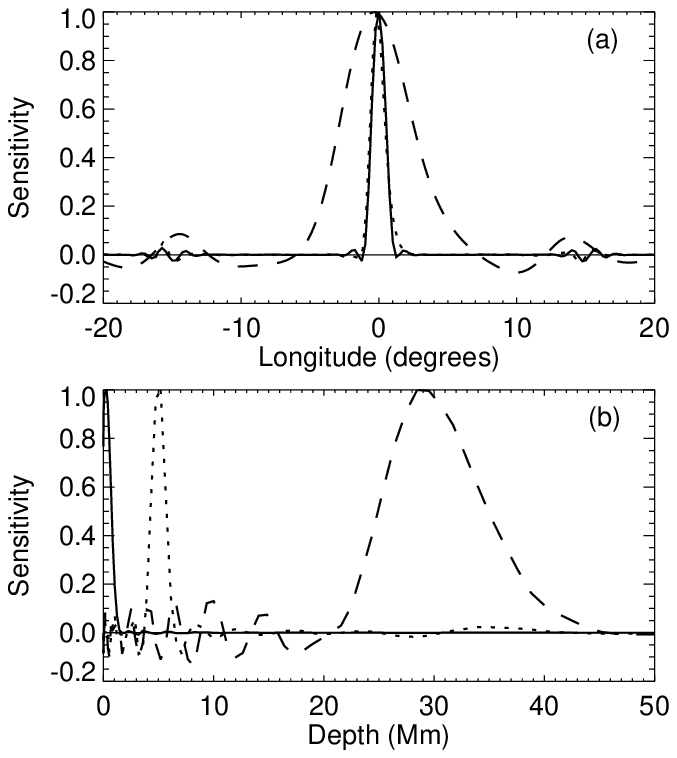}
	\caption{(a) Longitudinal cut through the peak of three different averaging kernels in the inversion, each targeting a different depth (Solid = 0.25 Mm, Dotted = 5 Mm, Dashed = 30 Mm). 
	(b) Cuts in depth of the same averaging kernels.
	Both the horizontal and vertical extent of the averaging kernels increase monotonically as a function of depth.
	}
	\label{fig:avgkers}
\end{figure*}

In Figure \ref{fig:avgkers}, we show cuts of three averaging kernels computed as part of the inversion. 
The three averaging kernels are for depths of $0.25 \ \mathrm{Mm}$, $5 \ \mathrm{Mm}$, and $30 \ \mathrm{Mm}$, spanning the depth range of our results. 
As shown in Figure \ref{fig:avgkers}a, the averaging kernels increase in horizontal size as the depth increases. 
Immediately below the surface, the inversion achieves a horizontal resolution of $6 \ \mathrm{Mm}$. 
The resolution scale increases with depth, degrading to $100 \ \mathrm{Mm}$ at a depth of $30 \ \mathrm{Mm}$.
The loss of horizontal resolution with depth is due to increasing uncertainty for Doppler shift measurements that reach deeply.
At all depths, the averaging kernels demonstrate the ability for the procedure presented here to resolve flows much smaller than the tile size. 
Figure \ref{fig:avgkers}b shows the vertical structure of the same three kernels. 
These give a sense of how much each depth in the solution is correlated with any other depth. 
The averaging kernels for all other depths presented in this paper show a steady progression of the vertical and horizontal extents as a function of depth. 


\section{RESULTS}
\label{sec:results}
\begin{figure*}
	\centering
	\includegraphics{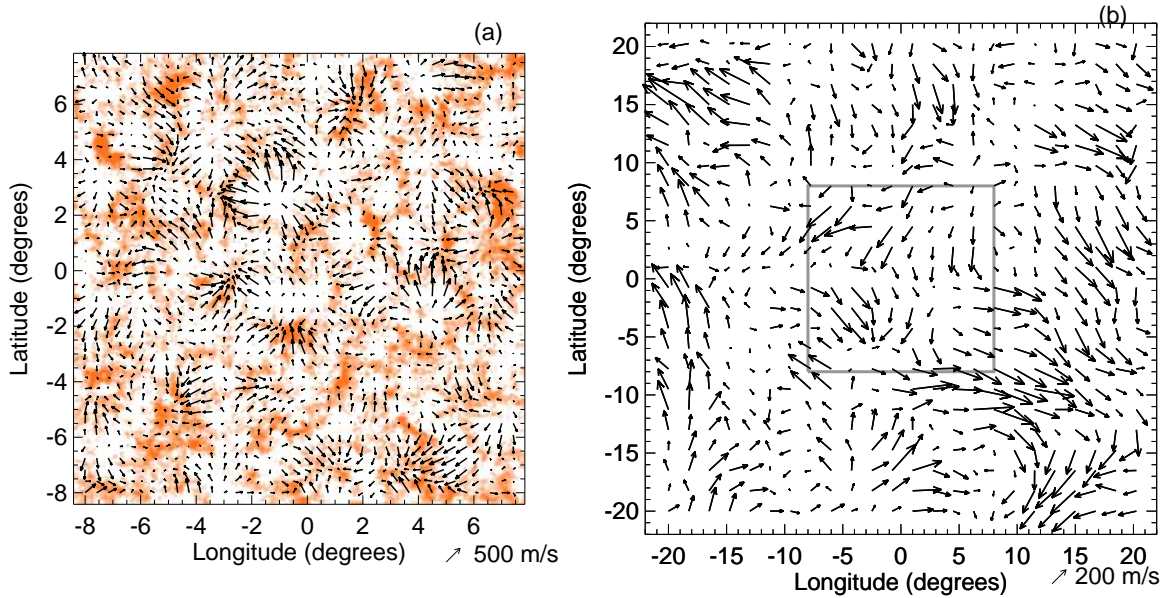}
	\caption{(a) Vector field for a subsection of the horizontal flow map at a depth of $0.25 \ \mathrm{Mm}$. 
	This map is extracted from disk center and is overlaid on a map of magnetic field strength. 
	Darker colors indicate stronger magnetic field, and the color table saturates at 25 Gauss.
	The vectors have been subsampled to a resolution of $0.5^{\circ}$ ($6 \ \mathrm{Mm}$) to match the width of the averaging kernel for this depth. 
	The magnetic field map shown is an average of the magnetic field strength (absolute value) as measured by HMI over the same time period that is averaged over in the velocity fields. 
	We see a strong correspondence between the horizontal flow found through our analysis and the advection of magnetic field at the photosphere. 
	(b) Vector field for a larger subsection of the inversion solution taken at a depth of $10 \ \mathrm{Mm}$ and subsampled to a resolution of $1.8^{\circ}$ ($22 \ \mathrm{Mm}$) to match the averaging kernel at this depth. 
	The gray box in panel (b) indicates the position and spatial extent of the map in panel (a).
	}
	\label{fig:flowmap}
\end{figure*}

Figure \ref{fig:flowmap}a shows a small region of the inversion solution at a depth of $0.25 \ \mathrm{Mm}$, and compares our helioseismically determined flow field to a map of the temporal average of the absolute, line-of-sight, magnetic field strength as measured by HMI over the same region. 
Magnetic field elements seen in photospheric observations are advected by the horizontal flows inside the Sun and tend to gather in the boundaries between neighboring convection cells. 
Averaged over the course of a day, the magnetic field map traces out the locations of supergranules, which persist for a comparable amount of time. 
Since the magnetic field map is an independent measure of the size and locations of supergranules, it serves as a useful comparison to our horizontal flow field. 
Figure \ref{fig:flowmap}a shows excellent agreement between the two, with collections of magnetic field sitting in the regions of converging flow. 
Figure \ref{fig:flowmap}b shows a larger portion of the inversion result at a depth of $10 \ \mathrm{Mm}$. 

In Figure \ref{fig:spectrum}, we present the velocity spectrum of the inversion results at four depths ($0.25 \ \mathrm{Mm}$, $5 \ \mathrm{Mm}$, $15 \ \mathrm{Mm}$, $30 \ \mathrm{Mm}$) as a function of spherical harmonic degree, and show the spectrum of the averaging kernels at each of these depths. 
Since the inversion flow field at a given depth is a horizontal convolution of the true flow field in the Sun with the averaging kernel at that depth, the spectrum of the averaging kernel provides the linear sensitivity of our procedure to solar flows at each harmonic degree. 
By construction, the averaging kernels integrate to unity, so the spectral sensitivity peaks at $\ell=0$ at a value of 1.
The sensitivity remains constant up to some value of $\ell$, then drops steadily thereafter. 
The location in spectral space at which this transition occurs is determined by the width of the averaging kernels (smaller width, larger $\ell$). 
We can therefore interpret the velocity spectrum at each depth as being reliable where the averaging kernel spectrum is flat, and artificially diminished for higher $\ell$.



At a depth of $0.25 \ \mathrm{Mm}$, the velocity spectrum shows a prominent bump between $\ell=50$ and $\ell=200$, consistent with the prominent supergranulation seen in Figure \ref{fig:flowmap}a. 
The location and amplitude of this feature is also consistent with measurements using other methods (e.g., \cite{2012A&A...540A..88R}, \cite{hathaway_2000}).
The averaging kernel spectrum at this depth confirms that we are able to resolve these scales, and that the steady drop in power above $\ell=200$ is due to limitations in resolution as opposed to an intrinsic drop in velocity at smaller scales. 
Deeper in the Sun, we lose the ability to sample supergranular scales as the horizontal resolution degrades. 
At a depth of $30 \ \mathrm{Mm}$, we are capable of resolving only the lowest harmonic degrees ($\ell < 40$).

Figure \ref{fig:rms} shows the root-mean-square (rms) horizontal velocity plotted as a function of averaging-kernel depth along with both the error on an individual solution point (shading) and the propagated error on the rms (error bars). 
The rms velocity peaks at $427 \ \mathrm{m \ s^{-1}}$ near the surface and diminishes rapidly with depth. 
Within the first $10 \ \mathrm{Mm}$ below the photosphere, the velocity drops to around $200 \ \mathrm{m \ s^{-1}}$ and continues a steady but slow decline until a depth of $30\ \mathrm{Mm}$. 
The small perturbations that appear within these deeper layers are within the error bounds and likely are a consequence of the specific realization of convective flows that we sample.


\section{DISCUSSION}
\label{sec:discussion}

\begin{figure*}
	\centering
	\includegraphics{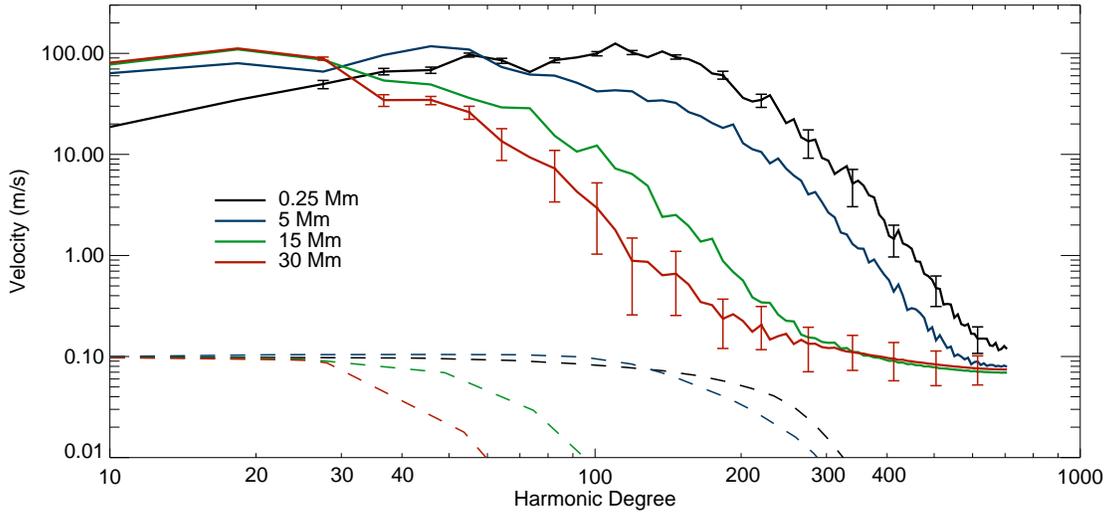}
	\caption{Spectrum of inversion results at a few depths as a function of harmonic degree. 
	Solid lines are the azimuthally integrated spectra of the horizontal velocity field, and color differentiates the central depth of the averaging kernel. 
	Vertical error bars on the black and red curves indicate the 95\% confidence interval, and those for the blue and green curves show similar trends. 
	The dashed lines are the upper envelope of the averaging kernels for each depth and demonstrate the relative sensitivity of our inversion result to the true flows in the Sun. 
	The sensitivity for every averaging kernel is unity at $\ell=0$ by definition, but the dashed lines have been normalized to arbitrary values for visual clarity.
	}
	\label{fig:spectrum}
\end{figure*}

\begin{figure*}
	\centering
	\includegraphics{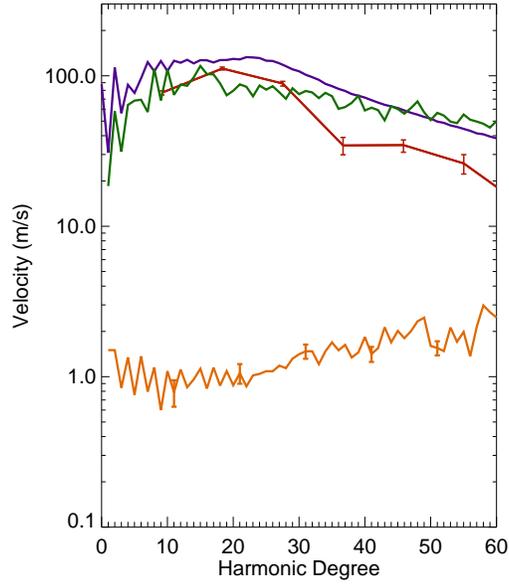}
	\caption{Comparison of horizontal velocity spectra as a function of harmonic degree.  
	The red curve indicates the inverted flow field (same as Figure \ref{fig:spectrum}), and error bars indicate the 95\% confidence interval at each value of harmonic degree. 
	The orange curve is the upper limit on convective amplitudes as appears in HDS12. 
	The purple curve is from the numerical hydrodynamic simulation described in Section \ref{sec:discussion}. 
	These three spectra are taken at a depth of approximately $30 \ \mathrm{Mm}$ ($0.96 \ \mathrm{R_{\odot}}$).
	The green curve is from the numerical simulation in \cite{miesch_2008} at a depth of $14 \ \mathrm{Mm}$ ($0.98 \ \mathrm{R_{\odot}}$).
	}
	\label{fig:compare}
\end{figure*}

Our primary finding is that the speed of solar convective flows exceeds $120 \ \mathrm{m \ s^{-1}}$ throughout the near-surface shear layer.  
This finding is in stark contradiction with the previous helioseismic study of HDS12. 
Figure \ref{fig:compare} compares the velocity spectra from Figure \ref{fig:spectrum} for low harmonic degree (red) with the time-distance helioseismic result of HDS12 (orange). 
The velocity spectrum for the numerical simulation of global convection used for comparison in HDS12 is shown in green (\cite{miesch_2008}). 
Further, the spectrum of motions from a more recent numerical simulation of the global convection zone is indicated (purple). 
This particular simulation evinces convective motion capable of sustaining a solar-like differential rotation, possessing a pole-to-equator contrast of $\Delta \Omega / \Omega \approx 15\%$. 
This model was computed by solving the anelastic equations in a rotating spherical shell using the numerical algorithms described in \cite{clune_1999} and a prescription for boundary conditions and radiative heating as described in \cite{featherstone_2014}.  
The simulation domain spans from the base of the convection zone at $0.72 R_\odot$ to a height of $0.97 R_\odot$ with a resolution of $128\times384\times768$~($n_r \times n_\theta \times n_\phi$).  
The two helioseismic spectra and the newer simulation spectrum are taken at a depth of approximately $30 \ \mathrm{Mm}$ ($0.96 \ \mathrm{R_{\odot}}$), and the spectrum from \cite{miesch_2008} is taken at a depth of $14 \ \mathrm{Mm}$ ($0.98 \ \mathrm{R_{\odot}}$).

Note, the time-distance study does not claim to have directly detected the convective flow signal.  
Instead, the indicated spectrum is an upper limit that depends on a specific noise model.  
The drastic difference between the two helioseismically determined spectra is particularly interesting, since each set of results use full-disk Dopplergrams from the same instrument. 
Research comparing results from time-distance analysis to those from ring-diagram analysis have generally shown good agreement (\cite{2011JPhCS.271a2005K}, \cite{hindman_2003}, \cite{hindman_2004}), and differences seen between the two techniques are usually far less substantial than what is shown in Figure \ref{fig:compare}.

\begin{figure*}
	\centering
	\includegraphics{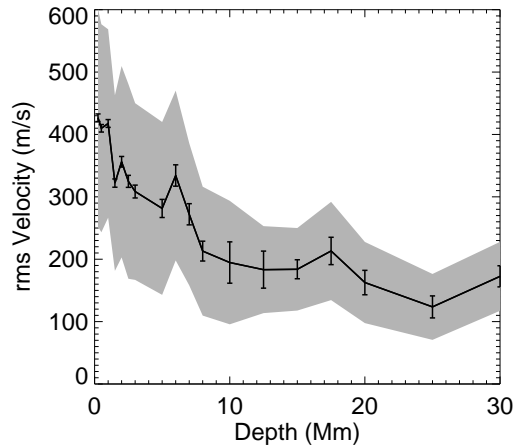}
	\caption{Root-mean-square horizontal velocity as a function of depth.
	The shaded region indicates the uncertainty on a single solution point while the vertical error bars indicate the 3-sigma values for the propagated uncertainty on the rms value. 
	}
	\label{fig:rms}
\end{figure*}

One of the key steps to the analysis in HDS12 is the attempted removal of uncorrelated noise in the variance of the velocity measurements. 
This is accomplished based on the assumption that the relevant flows produce a constant signal as the time duration of the analysis increases, while the level of uncorrelated noise decreases. 
If we were to assume that the discrepancy between the red and orange curves in Figure \ref{fig:compare} is simply due to the presence of noise in our data (which we have made no attempt to remove), the signal-to-noise ratio for the results presented would be at most $0.01$. 
However, by propagating the uncertainty measured directly from each power spectrum through the inversion procedure, we find our final signal-to-noise ratio to be around 2.5 (Figure \ref{fig:rms}). 
Perhaps, a more likely option is that the flow structures seen in this study evolve with time in such a way that they are removed during the noise-subtraction procedure of HDS12. 

Just as global helioseismic measurements of the subsurface differential rotation have guided numerical models of solar convection, our measurements of the convective amplitude provide useful observational constraints in the near-surface shear layer. 
The convective amplitude in the deepest layers that we sample (30 Mm) are particularly instructive, as these results are beginning to sample the deep flow structures responsible for the Sun's differential rotation and global meridional circulation. 
Such organized, large-scale motions are the result of Reynolds stresses induced by Coriolis deflection of the convective motions. 
The level of rotational influence felt by the convection, characterized by the Rossby number ${\rm Ro} = U/\Omega L$, is thus a crucial ingredient in simulations of the solar convection zone. 
Here $U$ is a typical velocity amplitude, $\Omega$ is the rotation rate, and $L$ is a characteristic length scale associated with the convection.  
Values of Ro greater than unity are characteristic of convection that only weakly senses the rotation, whereas values less than unity indicate rotationally constrained convection.  

A local Ro can be defined using a length scale corresponding to the peak in the velocity power spectrum (e.g., \cite{gastine_2013}). 
Based on our inversion data at 30 Mm, we choose a $U$ of 173 m s$^{-1}$ and a length scale of 292 Mm (corresponding to $\ell=15$). 
We deduce Ro at that depth to be 0.11.  
In the near-surface layers, we find Ro to be 2.16, using an $\ell$ of 120 and a $U$ of 427 m s$^{-1}$.  
Thus, based on our measured convective amplitude, the convective motions in the near-surface shear layer sense the Sun's rotation only weakly, but transition to a regime of high rotational constraint by 30 Mm. 

A more relevant Rossby number could be computed using the vertical flows, which our measurement procedure is insensitive to. 
However, one should be able to exploit the anelastic approximation to derive the vertical component using measured horizontal flows and mass conservation (\cite{komm_2004}).
Such an analysis is one of our future goals.


The analysis presented here provides not only a robust measure of the horizontal convective velocities in the solar near-surface shear layer, but also a proof-of-concept for further work in high-resolution ring-diagram helioseismology. 
While the computational burden of producing highly-overlapped tiles is significant, the ability to recover fine structures without sacrificing depth is of great importance. 
With larger tiles, it may be possible to extend this type of analysis deeper into the sun to sample the largest scales of convection with adequate resolution.  

We acknowledge D. O. Gough for making the initial suggestion that high-spatial resolution and deep sampling could be achieved through the deconvolution of many highly overlapped tiles of large size. 
We also thank S. Hanasoge for providing time-distance measurements and M. Miesch for providing simulation data, both illustrated in Figure \ref{fig:compare}.
This work was supported by NASA through NASA grants NNX09AB04G, NNX14AC05G, and NNX14AG05G. 
SDO is a NASA mission, and the HMI project is supported by NASA contract NAS5-02139.




\end{document}